\documentclass[prd,eqsecnum, showpacs,nofootinbib]{revtex4}
\global\arraycolsep=2pt
\usepackage{stmaryrd}
\usepackage{amsmath}
\usepackage{amssymb}
\usepackage{graphicx}
\usepackage{textcomp}
\usepackage{calrsfs}
\usepackage{yfonts}
\usepackage{bm}
\newcommand{\ben}{\begin{equation*}}                                       
\newcommand{\een}{\end{equation*}}
\newcommand{\bean}{\begin{eqnarray*}}                                   
\newcommand{\eean}{\end{eqnarray*}}

\newcommand{\be}{\begin{equation}}                                                  
\newcommand{\ee}{\end{equation}}
\newcommand{\bea}{\begin{eqnarray}}                                                 
\newcommand{\eea}{\end{eqnarray}}

\begin{document}

\title{Schwinger's Approach to Einstein's Gravity and Beyond\footnote{Based 
on invited
talk given at the American Astronomical Society Meeting, Anchorage, June 2012}}
\author{K. A. Milton}

\affiliation{Laboratoire Kastler Brossel, Universit\'e Pierre et Marie Curie,
Campus Jussieu Case 74, F-75252 Paris Cedex 05, France}
\altaffiliation{Permanent address:
H. L. Dodge Department of Physics and Astronomy,
University of Oklahoma, Norman, OK 73019 USA}

\begin{abstract}
Julian Schwinger (1918--1994), founder of renormalized quantum electrodynamics,
was arguably the leading theoretical physicist of the second half of the 20th
century.  Thus it is not surprising that 
he made contributions to gravity theory
as well.  His students made major impacts on the still uncompleted program
of constructing a quantum theory of gravity.  Schwinger himself had no
doubt of the validity of general relativity, although he preferred a
particle-physics viewpoint based on gravitons and the associated fields,
and not the geometrical picture of curved spacetime.  This note provides
a brief summary of his contributions and attitudes toward the subject of
gravity.
\end{abstract}

  
\maketitle

\section{Introduction}

 Julian Schwinger, founder, along with Richard Feynman and Sin-itiro
Tomonaga, of renormalized Quantum Electrodynamics in 1947-48, was the
first recipient (along with Kurt G\"odel) of the Einstein prize.
(For biographical information about Schwinger see Refs.~\cite{bio,update}.)
He was always deeply appreciative of Einstein's contributions to
relativity, quantum mechanics, and gravitation, and late in his career
wrote a popularization of special and general relativity called
{\it Einstein's Legacy} \cite{el}, based on an Open University course.

Thus it is surprising that at this
late date, almost 20 years after Schwinger's death, and nearly 60 after
Einstein's, to learn there was a scientific controversy between the two,
as expressed through an AAS session entitled ``Schwinger vs.\ Einstein.''

Schwinger in fact made major contributions to the development of the
quantum theory of gravity, based upon the Einstein equations, and
then went on to propose a source-theory formulation of the theory
of gravity, building on the notion that the carrier of the force of gravity
is the helicity-2 graviton, just as quantum electrodynamics 
is built on the hypothesis of
a helicity-1 photon.  From this starting point most of the consequences
of general relativity 
could be produced, including the classic tests of the redshift,
perihelion precession, the bending of light, and geodetic precession.
However, for strong fields, Schwinger showed that Einstein's full gravitational
field equations were a necessary consequence.

\section{Quantum Gravity}
The earliest example of a non-Abelian theory is gravity.  That is, the gauge
boson for gravity, 
the graviton, interacts directly with itself, unlike the photon in
electrodynamics.  So after writing several papers on non-Abelian theories
in the 1960s, Schwinger turned to gravity \cite{js65}.
These papers  made contact with the somewhat earlier work of two of his 
students, Richard Arnowitt and Stanley Deser \cite{adm}.  In his papers,
Schwinger introduces canonical variables, basically vierbeins (or tetrads)
and connections, and attempts to show quantum consistency.  Lorentz invariance
of the theory is verified subject to ``rather loosely stated physical boundary
conditions.''

\section{Source Theory of Gravity}
The complexity of these papers pushed Schwinger over the edge to 
{\it Source Theory.}  Source theory is a formulation of quantum field theory
in which Green's functions play a central role; in fact the basic objects
of any field theory are the Green's functions, which express all the physical
observables and correlations of the theory.  Explicit reference to
operator-valued fields is avoided. Green's functions and
sources always were always a vital part of his repertoire from the beginning of
his work in the 1930s, and in fact it is not a stretch to say that the 
 first ``source theory'' paper was his most famous paper, ``Gauge Invariance
and Vacuum Polarization,'' published in 1951 \cite{givp}.

Starting in 1968, Schwinger published several works on the source-theory
formulation of gravity \cite{sg,psf,pt,sp}.

\subsection{Graviton action}
The source of a massless, helicity-2 graviton is a conserved, symmetrical 
stress-tensor,
\be
\partial_\mu T^{\mu\nu}=0, \quad T^{\mu\nu}=T^{\nu\mu},
\ee
from which the generating function for all the Green's functions, 
the vacuum persistence amplitude, 
follows:
 \be
\langle 0_+|0_-\rangle^T=e^{iW[T]},\quad
 W[T]=\frac12\int (dx)(dx')\left[
 T^{\mu\nu}(x) D_+(x-x')T_{\mu\nu}(x')-\frac12 T(x)
D_+(x-x')T(x')\right],\label{action}\ee
which describes the free propagation of gravitons between sources.
Here appears the causal or Feynman massless propagator,
\be
D_+(x-x')=\int \frac{(dp)}{(2\pi)^4}\frac{e^{ip(x-x')}}{p^2-i\epsilon},
\ee
where  $p^2=-(p^0)^2+\mathbf{p\cdot p}$ and $T$ is the trace of the
graviton source tensor,
$T(x)=T^\mu{}_{\mu}(x).$  The above is expressed in natural units;
to connect to the real world, we rescale the source:
\be
 T^{\mu\nu}=\sqrt{\kappa}t^{\mu\nu},\quad \kappa=8\pi G,
\ee
where $G$ is Newton's constant.

Because gravity is of infinite range, the graviton should be massless.
However, Schwinger showed how you can start with a massive spin-2 particle,
of mass $m$, which has 5 helicity states.  Then provided we define
\be
\partial_\mu T^{\mu\nu}=\frac{m}{\sqrt{2}}J^\nu,\quad
\partial_\mu J^\mu=m\left(\sqrt{3}K-\frac1{\sqrt{2}}T\right),\label{decomp}
\ee
in the limit  $m\to 0$,
 $\partial_\mu T^{\mu\nu}=0$, $\partial_\mu J^\mu=0$,
and the action decouples into independent helicity 2, 1, and 0 components,
represented by a tensor source $T^{\mu\nu}$, a vector source $J^\mu$, and
a scalar source $K$.  As explained, for example, in Ref.~\cite{psf}, the 
coefficients in Eq.~(\ref{decomp}) allow for the decomposition into the three
helicity components, and the count of states between those for the massive
tensor description, and the massless helicity 2, 1, and 0 description is
$5=2+2+1$, suggesting a correspondence with the graviton, the photon,
and a massless scalar.

\subsection{Tests of General Relativity}
In the Physical Review paper \cite{sg} on the source theory of gravity, 
and in his
more detailed discussion in {\it Particles, Sources, and Fields} \cite{psf}, 
Schwinger rederives,  simply, just starting from the expression
for the graviton action given in Eq.~(\ref{action}), the standard tests of 
general relativity:
\begin{itemize}
\item The gravitational red shift,
\item the light deflection by the sun,
\item the time delay in radar echos from planets,
\item the precession of Mercury's perihelion.
\end{itemize}
These arguments, or ones very similar thereto, of course had been supplied
earlier by others.  For a bibliography of tests of gravity, see 
Ref.~\cite{Will:2010uh}.

In a couple of short articles in the American Journal of Physics
 a few years later (1974) \cite{pt,sp}, Schwinger
extended these elementary derivations to precession tests, which had not
been performed up to that time:
 The Thirring effect, the precessional
 angular velocity associated with
a rotating shell of radius $R$, mass $M$, and angular velocity $\bm{\omega}$:
\be
\bm{\omega}_{\rm prec}=\frac43 \frac{GM}R \bm{\omega},
\ee
and the Schiff effect, the precession 
(precessional velocity $\bm{\Omega}$) of a gyroscope in a satellite 
in orbit around a planet:
\be
\bm{\Omega}=\frac32\frac{GM}{r^3}\mathbf{r\times v}+\frac{GI}{r^5}(3\mathbf{r}
\bm{\omega}_p\cdot \mathbf{r}-\bm{\omega}_p r^2).
\ee
Here $r$ is the radius of the orbit, $\mathbf{v}$ is the velocity of the 
satellite, and $\bm{\omega}_p$ is the angular velocity
of the planet, which has moment of inertia $I$.
I followed these papers up with a simple source  theory rederivation of
the Lense-Thirring effect \cite{lt},  the effect
of the spin of the sun (mass $M$) 
on the motion of a planet (mass $m$) orbiting it, in terms
of the precession of its axial vector\footnote{Often referred to as
the Laplace-Runge-Lenz vector, or by some subset of those names,
but actually first discovered by Hermann in 1710 \cite{herman},
and generalized by Bernoulli in the same year \cite{bernoulli}.}
\be
\mathbf{A}=\frac{\mathbf{r}}r-\frac{M+m}{GM^2m^2}\mathbf{p\times L},
\ee
where $\mathbf{r}$ is the position of the planet relative  to the sun, 
$\mathbf{p}$ is the relative momentum, and $\mathbf{L}$ is the orbital angular
momentum.  The precession equation is
\be
 \frac{d\mathbf{A}}{dt}=\bm{\Omega}_A\times\mathbf{A},
\ee
where
\be
\bm{\Omega}_A=\frac{2G}{r^3}\left(\mathbf{S}-\frac{3\bm{\omega}_p(\bm{\omega}_p
\cdot \mathbf{S})}{\omega_p^2}\right),
\ee
where $\mathbf{S}$ is the spin of the sun, and $\bm{\omega}_p$ is the angular 
velocity of the planet.

These ``frame-dragging'' and ``geodetic'' 
 effects have now been confirmed by Gravity Probe B
\cite{gpb}, although the Lense-Thirring effect was earlier seen by the LAGEOS
experiment \cite{lageos}.

\subsection{But Einstein's General Relativity emerges}
The theory proposed by Schwinger to this point is nothing but 
linearized gravity, 
with some extrapolation to include gravity itself as a source of
energy. The symmetric tensor gravitational field $h_{\mu\nu}$ is related
to the metric tensor $g_{\mu\nu}$ of general relativity 
and the flat-space Minkowski metric $\eta_{\mu\nu}=\mbox{diag}(-1,1,1,1)$ by
\be
g_{\mu\nu}=\eta_{\mu\nu}+2h_{\mu\nu},
\ee
where $h_{\mu\nu}$ is regarded as a small perturbation.
(Schwinger inserted the factor of 2 to simplify the equations of motion for
$h_{\mu\nu}$.)
 But Schwinger was quite aware this was inadequate.  In his  paper 
\cite{sg},
and especially in the last section of Vol.~I of his book \cite{psf}, he 
recognizes that gravitational gauge invariance,
\be
h_{\mu\nu}\to h_{\mu\nu}+\partial_\mu \xi_\nu+\partial_\nu\xi_\mu,
\ee
where the vector field $\xi_\mu$ is arbitrary, is necessarily generalized
to general coordinate invariance, 
and in that way he was led inexorably to Einstein's equations,
\be
R_{\mu\nu}(x)=\kappa\left[t_{\mu\nu}(x)-\frac12 g_{\mu\nu}t(x)
\right],
\ee
in terms of the usual Riemann curvature tensor,
and to the Einstein-Hilbert action,
\be W=\int (dx) [\mathcal{L}_m+\mathcal{L}_g],\quad
2\kappa\mathcal{L}_g=\sqrt{-g(x)}g^{\mu\nu}(x) R_{\mu\nu}(x),
\ee
where $\mathcal{L}_m$ is the matter Lagrangian.

\subsection{Scalar-tensor gravity}
He did go on to notice that conformal symmetry is spoiled by this theory,
which motivated him to develop his own version of scalar-tensor gravity,
which for weak coupling agrees with Brans-Dicke theory \cite{bd}, 
but which could
have different consequences in the cosmological domain \cite{mng}.

\section{Supersymmetry and supergravity}
Later in the 1970s, Schwinger was chagrined that he had not come up with the
idea of {\it supersymmetry} \cite{Wess:1974tw}, 
since he had developed the multispinor formalism
that naturally allowed the treatment of all spins on the same footing
\cite{psf}.  He
invited his former student, Stanley Deser, for a private audience
with Schwinger's group of students, postdocs, and faculty, and we
were initiated into the mysteries of supersymmetry and supergravity
\cite{Deser:1976eh}. A paper by Schwinger giving a
simple rederivation of supersymmetric ideas followed \cite{multi}.
Bob Finkelstein, Luis Urrutia,  
and I immediately continued with a paper in which we showed
that supergravity emerged by requiring solely invariance under local
supersymmetry \cite{superg}.

\section{Schwinger's attitude toward General Relativity}
 Schwinger certainly had no quibble with general relativity,
and provided an alternative derivation of Einstein's equations. In this
he behaved analogously to Richard Feynman who, in his Polish lecture, also
provided a particle physics derivation of general relativity
\cite{Feynman:1996kb}.  In fact, Schwinger privately stated that he
thought he would have discovered general relativity had he been in Einstein's 
place.

It is true that Schwinger's approach was always algebraic,
and consequently
he had little use for the geometric interpretation of curved space. Yet
this is merely an interpretation of the theory, and
had no effect on testable consequences.
 Schwinger did not believe in a massive graviton, and certainly
not in tachyons.

\section{Schwinger's  attitude toward the unseen universe}

 Schwinger never expressed an opinion on dark matter, to my knowledge.  
In those
days, dark matter, manifested by galactic rotation curves,
 was largely only of interest to astronomers, and not
to physicists in general \cite{Bertone:2004pz}.
 Of course, he didn't know about ``cosmic acceleration'' 
\cite{Frieman:2008sn}, so
he likely thought that the cosmological constant was zero.
Since he became very fascinated with the Casimir effect \cite{lmp},
I'd like to imagine he would have thought that dark energy 
originated from a nonzero
cosmological constant, arising from quantum fluctuations.
Such ideas go back at least to Pauli \cite{pauli}, and for a history
of the connection of the ideas of zero-point energy and the cosmological
constant see Kragh \cite{kragh}.  Particularly noteworthy are the contributions
of Gliner \cite{gliner}, Zel'dovich \cite{zeldovich}, and Sakharov 
\cite{sakharov}.  See also Ref.~\cite{birrell}.

An idea along these lines was proposed some time ago \cite{milton01}.
For example, quantum fluctuations of fields in ``large'' 
compactified dimensions 
would give rise to a cosmological constant: ($a=$ size of compact space
of dimension $d$)
\be
\langle T^{\mu\nu}\rangle
=-u g^{\mu\nu}=-\frac\Lambda{8\pi G}g^{\mu\nu}.
\ee

Roughly speaking, the quantum vacuum energy of the fluctuations must have the 
form
\be u=\frac{\gamma_d}{a^4},\quad d \mbox{\, odd},\qquad u=\frac{\alpha_d
\ln a/L_{\rm Pl}}{a^4}, \quad d \mbox{\, even}.\ee
for odd and even compactified dimensions $d$, respectively, where $L_{\rm Pl}$
is the Planck length.  The coefficients
$\alpha_d$ and $\gamma_d$ depend on the fields compactified \cite{kan}.
We must require that the density of dark energy be less than the critical 
density that would close the universe, which for a reduced Hubble
constant of $h_0=0.7$ corresponds to a length scale of 80 $\mu$m,
which leads to
\be a\ge\gamma^{1/4} 80 \,\mu\mbox{m}, \quad
a\ge [\alpha\ln(a/L_{\rm Pl})]^{1/4}80\,\mu\mbox{m},
\ee
which are nearly  excluded by the E\"otWash experiment \cite{eotwash},
\be
a\le 44 \mu \mbox{m}.\ee
See Table \ref{tab}, which is reproduced from Ref.~\cite{milton01}.
 \begin{table}
\centering
\begin{tabular}{lcccc}
$\cal S$&Gravity&Scalar&Fermion&Vector\\
\hline
$S^1$ (u)&*&*&9.5 $\mu$m&---\\
$S^1$ (t)&9.9 $\mu$m&6.6 $\mu$m&*&---\\
$S^2$&84 $\mu$m&*&*&*\\
$S^3$&---&7.5 $\mu$m&9.5 $\mu$m&---\\
$S^4$&*&*&*&77 $\mu$m\\
$S^5$&---&11.5 $\mu$m&*&---\\
$S^6$&350 $\mu$m&*&*&110 $\mu$m\\
$S^7$&---&13.5 $\mu$m&7.0 $\mu$m&---\\
\end{tabular}
\caption{\label{tab}
The lower limit to the radius of the compact dimensions
deduced from the requirement that the Casimir energy not exceed
the critical density. The numbers shown are for a single species of
the field type indicated.  The dashes indicate cases where the Casimir
energy has not been calculated, while
 asterisks indicate (phenomenologically
excluded) cases where the Casimir energy is negative.}
\end{table}
These ideas have beeb more recently elaborated, for example, in
Refs.~\cite{Greene:2007xu} and \cite{Dupays:2013nm}.

\section{Conclusions}

 Although Schwinger approached gravity from a particle-physics
viewpoint, he never expressed any doubt about the validity of general
relativity.   He thought algebraically, not geometrically, so he didn't find
the notions of curved space useful.
 Many of his students have made major contributions to the
quantum theory of gravity (for example,
beside Arnowitt and Deser mentioned above, one cannot forget
the contributions of Bryce DeWitt \cite{dewitt} or of David Boulware
\cite{Boulware:1985wk}), 
which, although still not fully developed \cite{Ashtekar:1987gu}, must reduce
to Einstein's theory under most circumstances.

\acknowledgments
I thank the Laboratoire Kastler Brossel for their hospitality, particularly
Astrid Lambrecht and Serge Reynaud.  CNRS is thanked for their support.
This work was further supported in part by a grant from the Simons Foundation;
earlier work summarized here was supported by grants from
the US Department of Energy.  I thank Ron Kantowski and
Yun Wang for earlier collaborations
on the dark energy scheme.  I thanks Stanley Deser for critical comments on
the first version of this paper.

\end{document}